\documentstyle[12pt,epsf,eepic,axodraw]{article}

\newcommand{\be}{\begin{equation}}
\newcommand{\ee}{\end{equation}}
\newcommand{\bea}{\begin{eqnarray}}
\newcommand{\eea}{\end{eqnarray}}
\newcommand{\eps}{\epsilon}

\newcommand{\aha}{{\hat{a}}}
\newcommand{\J }{{\cal J}}
\newcommand{\I }{{\cal E}}
\def\Wslash{\not{\hbox{\kern-4pt $W$}}}
\def\Qslash{\not{\hbox{\kern-4pt $Q$}}}
\def\eslash{\not{\hbox{\kern-2pt $\varepsilon$}}}
\def\pslash{\not{\hbox{\kern-2.3pt $p$}}}
\def\kslash{\not{\hbox{\kern-2.3pt $k$}}}
\def\qslash{\not{\hbox{\kern-2.3pt $q$}}}

\def \L{{\widehat{L}}}

\def \tr {\rm {tr}}
\def\dslash{\not{\hbox{\kern-2.4pt $\partial$}}}

\begin{document}
\begin{flushright}
\hfill{SLAC-PUB-8184}\\
\hfill{SU-ITP-99/32}\\
\hfill{hep-th/9906239}\\
\hfill{\today }\\
\end{flushright}

\vspace{20pt}

\begin{center}
{\large {\bf One-loop corrections to the D3 brane action.}}

\vspace{40pt}

{\bf Marina Shmakova }

\vspace{20pt}
{\it 
Stanford Linear Accelerator Center,

Stanford University, Stanford, CA 94309;

University of Tennessee, Knoxville, TN 37996.}

\vspace{60pt}

\underline{ABSTRACT}

\end{center}

 We calculate one-loop corrections to the effective Lagrangian for the D3 brane. We perform the gauge-fixing of the $\kappa$-symmetric Born-Infeld D3 brane action in the flat background using Killing gauge. The linearized supersymmetry of the gauge-fixed action  coincides with that of the ${\cal N}=4$ Yang-Mills theory. We use the helicity amplitude and unitarity technique to calculate the one-loop amplitudes at order $\alpha ^4$.  The counterterms and the finite 1-loop corrections are  of the form $(\partial F)^4$ and their supersymmetric generalization. This is to be contrasted with  the  Born-Infeld action which contains  $(F)^4$ and other terms which do not depend on derivatives of the vector field strength.

{\vfill\leftline{}\vfill
\vskip  30pt
\footnoterule
\noindent
{\footnotesize
 e-mail: shmakova@slac.stanford.edu. }  \vskip  -5pt

%\vskip  14pt

%\baselineskip=24pt
\pagebreak
\setcounter{page}{1}

\section{Introduction}

 The action for supersymmetric D3-branes in flat and curved type IIB
supergravity backgrounds has been studied extensively during the  
last few years \cite{Cederwal, BT, Schwarz, Tseytd3}.  In addition to extended supersymmetry these actions have a local $\kappa $-symmetry and  half of the 32  fermionic fields can be ``gauged away'' by fixing  the $\kappa $-gauge. Moreover, of the 32 supersymmetries of type IIB supergravity, 16 are  
realized linearly and  16 non-linearly. The interpretation of all these supersymmetries is closely related to the problem of the  correspondence between supersymmetric ${\cal N}=4$  $U(N)$-symmetric  Yang-Mills theories and D3-brane Born-Infeld-type $U(1)$-symmetric  actions \cite{Tseytd3}. The ${\cal N}=4$ Yang-Mills theory in d=4 is not only renormalizable but even finite \cite{Mandelstam}. The D3 brane action has been considered  so far only as an effective action as it is not renormalizable by power counting in d=4 . Still, there are 16+16 linear and non-linear global symmetries in the gauge-fixed action (or equivalently a local $\kappa$-symmetry of the classical action). Is it possible that these symmetries are strong enough to forbid the  counterterms in the D3  brane action which have a higher number of derivatives than the classical Born-Infeld action? One has to take  into account that the D3 brane action has terms of the form $\bigl((F_{\mu\nu})^n +\dots \bigr)$ however, terms of the form $(\partial F)^n $ or $(\partial^2  F)^n $ are not present.

As a first step in  resolving these issues, in particular to clarify  the relation between the effective action of the open string theory, Yang-Mills theory and effective action of the D3-brane theory,  we will treat the $\kappa$-symmetric action of the D3-brane on the same footing as supergravity in d=4 or supersymmetric Yang-Mills theory in $d> 4$. A priori these theories are not renormalizable. However a careful analysis of supersymmetric higher derivative counterterms was performed  in the past, as well as some actual calculations  \cite{MarcusSt}.  

In this paper we will present one-loop corrections to the D3-brane Abelian
Born-Infeld action in a flat background, and we will show that there are
ultraviolet divergences, in contrast to the finite  d=4 ${\cal N}=4$
Super-Yang-Mills  \cite{Mandelstam}, and that the form of the  couterterm is $(s^2+t^2+u^2)F^4 \sim (\partial F)^4 $. To fix the $\kappa $-symmetry we  will employ the so-called Killing gauge used in \cite{RenJa} for gauge-fixing of the GS type IIB string in the $ {\rm AdS}_5\times{\rm S}^5$ background. In this gauge we will be able to directly identify the linear supersymmetries of the gauge-fixed D3 brane action  with the YM supersymmetries. We will compare this to the symmetries of the D3 brane action in $\theta_1=0$ gauge, proposed in \cite{Schwarz}.  The calculation of the tree amplitudes and one-loop corrections gives the same results in the other gauge, of course. However, in the Killing gauge the lowest order Lagrangian contains only quartic interaction terms while in $\theta_1=0$ gauge there are cubic as well as quartic interactions. Therefore it is easier to perform one-loop calculations in the Killing gauge.

It is interesting  that the terms  with the $(\partial F)^4$  structure were found in the effective action for the open superstrings in the paper by Andreev and Tseytlin \cite{AndrTsey} from the string S-matrix term $(s^2+t^2+u^2)F^4.$  This kind of terms were also found in the more recent papers by Hashimoto and Klebanov \cite{HasKleb} in which they computed the scattering amplitude for massless vectors on a D-brane in string theory. In the low energy limit their result also contain the term proportional to $(s^2+t^2+u^2)F^4.$  The appearance of this structure in  our calculations and in the string effective action could  mean  that the supersymmetry fixes this structure in the unique way. This would also mean that the local $\kappa $-symmetry after gauge fixing corresponds to the global ${\cal N}=4$ supersymmetry at least in this approximation.

To perform one-loop calculations we will use the helicity amplitude method that has been  developed extensively in the last 10 years (for reviews see \cite{SHform,Onelooprev}).  The main ideas of this method  that we will employ here are the spinor helicity representation for polarization vectors \cite{Oneloop1}, supersymmetry identities \cite{susyamp} and unitarity cuts \cite{Oneloop4}. This method has been used for loop calculations in Yang-Mills theory  and recently also in gravity \cite{Maxim}. It turns out that it is extremely useful for the D3 brane calculations.

In Sec. 2 we will discuss the  D3-brane action  and find the SUSY transformations in the  Killing gauge. In Sec. 3 the  ${\cal N}=4$ supersymmetric Lagrangian of the Born-Infeld theory will be presented  up to the quartic order in the fields. In Sec. 4 we will apply the helicity amplitude formalism to the D3-brane action. We will present tree and one-loop amplitudes for the quartic Lagrangian in the Killing gauge and give the expressions for the one-loop corrections and counterterms in this theory. In  Sec. 5 we will discuss the form of the Lagrangian up to the quartic order  in  $\theta_1=0$ gauge and Sec. 6 is a conclusion.

   \section{ The  D3 brane action in a flat background}

In this paper we will closely follow the notations of \cite{Tseytd3}.  The supersymmetric D3-brane action  is a sum of the Born-Infeld (BI) and Wess-Zumino (WZ) terms \cite{Cederwal, Schwarz, Tseytd3} that depend on superspace coordinates $X^M=(x^{\hat a},\Theta ^I)$ and an Abelian world-volume gauge field strength  $dA$
  \begin{eqnarray}
S = S_{BI} + S_{WZ}\ ,
\label{action}
\end{eqnarray}
where the Born-Infeld part is
\begin{eqnarray}
S_{BI} =-T_3\int_{M_4}  d^4\sigma\  \sqrt{-\det{(G_{ij}+2\pi\alpha ^\prime {\cal F}_{ij})}} \, 
\label{action1}
\end{eqnarray}
where $T_3 \sim \frac{1}{(2\pi\alpha ^\prime)^2}$ is the D3-brane  tension \cite{Strings} and $2\pi\alpha ^\prime$ is the inverse string tension. $G_{ij}$ is the pullback to the d=4 world-volume metric of the d=10 Minkowski metric $\eta^{{\hat{a}}{\hat{b}}}=(-+...+)$  with $i,j=0...3 ;\,{\hat{a}},{\hat{b}}=0...9 :$ 
\begin{equation}
G_{ij}=L_i^{\hat{a}}\eta^{{\hat{a}}{\hat{b}}} L_i^{\hat{b}}
\,,  \qquad 
L^{\hat{a}} (X(\sigma)) = d\sigma^i L_i^{\hat{a}}\,
\end{equation}
 and in the flat background the supervielbeins are
 \bea L^{\cal I} &=& d\Theta^{\cal I},\quad  \quad {\cal I} =1,2\nonumber \\
 L^{\hat a} &=& dx^{\hat a} - {\rm i}\bar\Theta^{\cal I}\Gamma^{\hat a} d\Theta^{\cal I}. \label{svn1}
\eea
 The field strength ${F}=\frac{1}{2}{\cal F}_{ij}d\sigma^i\wedge d\sigma^j$ is the supersymmetric extension of $dA$: 
\begin{equation}\label{fexp} 
F=dA +2{\rm i}\int_0^1 dt\  L_t^{\hat{a}}\wedge  
\bar{\Theta}\eta^{ab}\Gamma^{\hat{b}} {\cal K} L_t \  . 
\label{acF}
 \end{equation}
 The WZ part of the action is  
\be
S_{WZ} = 2{\rm i} T_3 \int_{M_4}  \int_0^1 dt
 \,
\bigg(\frac{1}{6} \bar{\Theta}\widehat{L}_t\wedge \widehat{L}_t\wedge \widehat{L}_t {\cal E} L_t
+2\pi\alpha ^\prime \bar{\Theta}\widehat{L}_t\wedge{\rm F}_t\wedge {\cal J}L_t\bigg) + T_3 \int_{M_4} \Omega _4^{(bos)} \, ,
\label{acWZ}
\ee
where  $\widehat{L} = L^{\hat a}\Gamma^\aha $ and $ L_t^{\hat{a}}(x,\Theta)\equiv L^{\hat{a}}(x,t\Theta),\ \ L_t(x,\Theta)\equiv L(x,t\Theta).$ The $\Theta$ independent part of the 4-form in the flat background is $\Omega _4^{(bos)} = d^4\sigma.$ The $\Gamma^\aha$ are the d=10 Dirac matrices  and the  matrices ${\cal E},\J ,{\cal K}$  in  (\ref{acWZ}) and (\ref{acF}) act on the SO(2) indices of type IIB spinors $\Theta ^{\cal I}$:
\be
{\cal E}=\left(
\begin{array}{cc}
0  &  1\\
-1 &  0
\end{array}
\right)\,,
\qquad 
\J=\left(
\begin{array}{cc}
0  &  1\\
1   &  0
\end{array}
\right)\,,
\qquad
{\cal K}=\left(
\begin{array}{cc}
1   &  0\\
0   &   -1 
\end{array}
\right)
\,. \label{noon}\ee

The supersymmetry transformations with global parameter $\varepsilon $ and $\kappa $-symmetry transformations with local parameter  $\kappa (\sigma )$ for this action are \cite{Cederwal, Tseytd3}
\bea
\delta_\varepsilon \Theta ^{\cal I}&=&\varepsilon^{\cal I} \\
\delta_\varepsilon x^\aha  &=&-{\rm i}( \bar \Theta\Gamma^{\aha}\varepsilon )\\
\delta_\varepsilon A &=& {\rm i} d x^\aha \bar \Theta\Gamma^{\aha}{\cal K}\varepsilon -\frac{1}{6}(d\bar \Theta\Gamma_{\aha}\Theta)(\bar \Theta\Gamma^{\aha}{\cal K}\varepsilon ) -\frac{1}{6}(d\bar \Theta\Gamma_{\aha}{\cal K}\Theta)(\bar \Theta\Gamma^{\aha}\varepsilon )
\label{SUSYn}
\eea
\bea
\delta_\kappa \Theta ^{\cal I} &=&\kappa ^{\cal I}\\
\delta_\kappa x^\aha &=& {\rm i}( \bar \Theta\Gamma^{\aha}\kappa )\\
\delta_\kappa A &=& -{\rm i} d x^\aha \bar \Theta\Gamma^{\aha}{\cal K}\kappa +\frac{1}{2}(d\bar \Theta\Gamma_{\aha}\Theta)(\bar \Theta\Gamma^{\aha}{\cal K}\kappa ) +\frac{1}{2}(d\bar \Theta\Gamma_{\aha}{\cal K}\Theta)(\bar \Theta\Gamma^{\aha}\kappa ),
\label{kapn}
\eea
where the SUSY parameter $\varepsilon^{\cal I} $ is a constant type IIB Majorana-Weyl spinor. Half of the components of the $\kappa$-transformation parameter are  projected out by the condition
\be \label{kappan}
\kappa=\frac{1}{2}(1+\Gamma )\kappa \,, \qquad \ \ 
\Gamma^2=1,
 \ee
with
\bea
\Gamma = \left(
\begin{array}{cc}
0  &  \zeta\\
\tilde{\zeta} &  0
\end{array}\right)=\frac{\epsilon^{i_1\ldots i_4}\bigg(\frac{1}{4!} \Gamma_{i_1\ldots i_4} \I
+\frac{2\pi\alpha ^\prime}{4}\Gamma_{i_1i_2} {\cal F}_{i_3i_4} \J
+\frac{(2\pi\alpha ^\prime)^2}{8} {\cal F}_{i_1i_2}{\cal F}_{i_3i_4} \I\bigg)}{\sqrt{-\det(G_{ij}+2\pi\alpha ^\prime {\cal F}_{ij})}}
\label{ganzn}
\eea
where $\Gamma_{i_1\ldots i_n}\equiv \L_{[i_1}\ldots \L_{i_n]}.$

	Following \cite{RenJa, RenWVS} we will choose the ``Killing gauge'' for the $\kappa$-symmetry.  In \cite{RenJa} $\Theta^{1,2}$ were choose to be the Killing spinors of the $AdS_5$ background, for the flat background we can simply introduce new variables $\vartheta^{\pm}$
\be
\vartheta^{\pm}= {\it U}\Theta^{1,2}\quad \Rightarrow \quad 
\left(
\begin{array}{cc}
 \vartheta^{+} \\
\vartheta^{-}
\end{array}\right)=\frac{1}{\sqrt 2}\left(
\begin{array}{cc}
1  &  \Gamma^{(4)}\\
1 &  -\Gamma^{(4)}
\end{array}\right)\left(\begin{array}{cc}
 \Theta^{1} \\
\Theta^{2}
\end{array}\right)
\label{ngn}
\ee
where $\Gamma^{(4)}=\Gamma^{0}\Gamma^{1}\Gamma^{2}\Gamma^{3}$ with 
\bea
\Gamma^{p}&=&\gamma^p\otimes 1,\quad \, p=0,\ldots,3  \nonumber \\
\Gamma^{t}&=&\gamma^{5}\otimes \tilde{\gamma}^t,\quad \, t=4,\ldots ,9 
\eea
Here $\gamma^p$, and $\gamma^{5}$ are standard Dirac matrices with $\left\{\gamma^p,\gamma^{p^\prime}  \right\}=2\eta^{pp^\prime}$ and  $\tilde{\gamma}^t$  are $8\times 8$ six-dimensional Dirac matrices with 
$ \left\{ \tilde{\gamma}^ {t},\tilde{\gamma}^{t^\prime} \right\}= 2\delta^{tt^\prime}.$

The supersymmetry and $\kappa$-symmetry transformations  (\ref{SUSYn}) and (\ref{kapn}) with the  new parameters  ${\varepsilon}^{\pm}={\it U}\varepsilon^{\cal I}$,
 ${\kappa}^{\pm}={\it U}\kappa^{\cal I}$  are
\be
\left(
\begin{array}{cc}
 \delta_{\kappa+\varepsilon} \vartheta^{+} \\
\delta_{\kappa+\varepsilon} \vartheta^{-}
\end{array}\right)=\left( \begin{array}{cc}
 \varepsilon^{+}\\
\varepsilon^{-}
\end{array}\right)+\frac{1}{2}\left(
\begin{array}{cc}
1-\frac{1}{2}C  & \frac{1}{2}A\\
-\frac{1}{2}A & 1+\frac{1}{2}C
\end{array}\right)\left(\begin{array}{cc}
 \kappa^{+} \\
 \kappa^{-}
\end{array}\right).
\label{ksusyn}
\ee
Here the  matrices $A$ and $C$ are 
\be
 A =\zeta \Gamma^{(4)}+\Gamma^{(4)}\tilde{\zeta},\quad C=\zeta\Gamma^{(4)}-\Gamma^{(4)}\tilde{\zeta}
\ee
and explicit expressions for  $\zeta$ and $\tilde{\zeta}$ are given in (\ref{ganzn}).

We gauge away half of the fermionic coordinates  by imposing  $\vartheta^-=0.$ This condition also requires that  $\delta_{\kappa+\varepsilon} \vartheta^{-}=0.$  Gauge fixing  $\kappa$-symmetry by the condition $\kappa^{+}=0$ we get:
\bea
\delta_{\kappa+\varepsilon} \vartheta^{+}&=&\frac{1}{4} A \kappa^{-}+\varepsilon^{+}\label{eta+}\\
\delta_{\kappa+\varepsilon} \vartheta^{-}&=&\frac{1}{2}( 1+\frac{1}{2}C )\kappa^{-}+\varepsilon^{-}=0.\label{eta-}
\eea
Eq. (\ref{eta-}) gives a relation between $\kappa^{-}$ and  $\varepsilon^{-},$
\be
\kappa^{-}= -2 ( 1+\frac{1}{2}C )^{-1}\varepsilon^{-}.
\label{keps}
\ee
Finally we get 
\be
\delta_{\kappa+\varepsilon} \vartheta^{+}=\frac{1-\zeta\Gamma^{(4)}}{1+\zeta\Gamma^{(4)}}\varepsilon^{-}+\varepsilon^{+}.\label{susyeta+}\\
\ee
Using this transformation on $A^p$ and $x^{\hat a} = {x^p,\,y^t}$ ($p=0,\cdots , 3;$ $t=4,\cdots , 9$ ):
\bea
\delta_{\kappa+\varepsilon} x^p &=&i\bar\vartheta^{+}\Gamma^p(\tau\varepsilon^{-}-\varepsilon^{+})\\
\delta_{\kappa+\varepsilon} y^t &=&-2i\bar\vartheta^{+}\Gamma^t\varepsilon^{-}\\
\delta_{\kappa+\varepsilon} A^p &=&2i\bar\vartheta^{+}\Gamma^p\varepsilon^{-}-\frac{2}{3}\partial_{p} \bar\vartheta^{+} \Gamma_a \vartheta^{+}\bar\vartheta^{+} \Gamma^a \varepsilon^{-}-i\partial_{p} y^t \bar\vartheta^{+} \Gamma^t (\tau \varepsilon^{-} -\varepsilon^{+})
\eea
where 
\be
\tau = \frac{1-\zeta \Gamma^{(4)}}{1+\zeta \Gamma^{(4)}}
\label{tau1}.
\ee
 The additional general coordinate reparametrization is necessary  to satisfy the  static gauge $( x^p=\sigma ^p)$ condition $ \delta x^p= 0$ 
\bea
\delta_\xi \vartheta &=&\xi^p\partial_{p} \vartheta \\
\delta_\xi x^\aha &=&\xi^p\partial_{p}x^\aha 
\eea
then from $ \delta _{\kappa+\varepsilon +\xi }x^p= 0 $  and $\partial_{i}x^p = \delta _i^p$  it follows that $\xi^i= - i\bar\vartheta^{+}\Gamma^i(\tau\varepsilon^{-}-\varepsilon^{+}).$
Finally the SUSY-transformations for  remaining nonzero fermions, scalars and vectors are:
\bea
\delta_{\kappa+\varepsilon+\xi} \vartheta^{+}&=&(\tau \varepsilon^{-} + \varepsilon^{+}) +\xi^i\partial_{i} \vartheta^{+} \\
\delta_{\kappa+\varepsilon+\xi} y^t &=&-2i\bar\vartheta^{+}\Gamma^t\varepsilon^{-} + \xi^i\partial_{i}y^t \\
\delta_{\kappa+\varepsilon+\xi} A^i &=&2i\bar\vartheta^{+}\Gamma^i\varepsilon^{-}-\frac{2}{3}\partial_{i} \bar\vartheta^{+} \Gamma_a \vartheta^{+}\bar\vartheta^{+} \Gamma^a \varepsilon^{-}\\
&-&i\partial_{i} y^t \bar\vartheta^{+} \Gamma^t (\tau \varepsilon^{-}-\varepsilon^{+})+(\xi^j\partial_{j}A^i+\partial_{i} \xi^j A^j) 
\eea
It is easy to see that the  transformations for $y^t$ are linear and do not contain an  $\varepsilon^{+}$ part.

     From here on  we will drop  the ``$+$''  in the spinor notation i.e. $\vartheta^{+} \Rightarrow \vartheta .$ In the new variables the  supervierbeins (\ref{svn1}) are:
\bea
L=d\vartheta ;  \quad L^{p}=dx^{p} - i\bar\vartheta \Gamma^{p} d\vartheta ;\quad L^{t}= dy^{t}.\label{La2} 
\end{eqnarray}

\section{ The Quartic ${\cal N}=4$  Lagrangian}

 The D3-brane BI action (\ref{action1}) simplifies considerably in the Killing gauge. In this case 
\begin{eqnarray}
G_{ij}+2\pi\alpha ^\prime {\cal F}_{ij} &= &\eta^{ij}-i\bar\vartheta \Gamma_i\partial_{j}\vartheta -i\bar\vartheta \Gamma_j\partial_{i}\vartheta -\bar\vartheta \Gamma_p\partial_{i}\vartheta \bar\vartheta \Gamma_p\partial_{j}\vartheta +\partial_{i}y^t \partial_{i}y^t \nonumber \\
&+ &2\pi\alpha ^\prime ({\rm F}_{ij}-i\partial_{i}y^t\bar\vartheta \Gamma_p\partial_{j}\vartheta + i\partial_{j}y^t\bar\vartheta \Gamma_t\partial_{i}\vartheta)  
\label{action3}
\end{eqnarray}
where ${\rm F}_{ij}=\partial_{i}A_j-\partial_{j}A_i$ and the WZ-term is 
\bea
{\it L}_{WZ}&=& -T_3\bigl( i\bar \theta\dslash \theta -\frac{i}{2}\partial_l \bar \theta \Gamma^{jkl}\Gamma^{t}\Gamma^{t^\prime}\theta \partial_j y^ {t}\partial_k y^ {t^\prime} + \frac{1}{2}({(\bar \theta\dslash \theta )}^2-\bar \theta \Gamma^{i}\partial_j\theta \bar \theta \Gamma^{j}\partial_i\theta ) \nonumber \\
&-& \frac{1}{4}\epsilon^{ijkl}\partial_k\bar\theta \Gamma^{c}\theta \partial_l\bar\theta \Gamma_{c} \Gamma^{(4)}\Gamma^{t}\Gamma^{t^\prime}\theta \partial_i y^ {t}\partial_j y^ {t^\prime}-\frac{1}{2}\epsilon^{ijkl}\partial_k \bar \theta \Gamma^{t}\Gamma^{(4)}\theta \partial_l\bar\theta \Gamma^{t^\prime}\theta \partial_i y^ {t}\partial_j y^{t^\prime}\nonumber \\
&+& \frac{i}{6}\epsilon^{ijkl}\epsilon_{abcl}\bar \theta \Gamma^{a}\partial_i\theta \bar \theta \Gamma^{b}\partial_j\theta \bar \theta \Gamma^{c}\partial_k\theta + \frac{1}{24}\epsilon^{ijkl}\epsilon_{abcd}\bar \theta \Gamma^{a}\partial_i\theta \bar \theta \Gamma^{b}\partial_j\theta \bar \theta \Gamma^{c}\partial_k\theta \bar \theta \Gamma^{d}\partial_l\theta   \nonumber \\
&- &(2\pi\alpha ^\prime)\frac{i}{2}\partial_l \bar \theta \Gamma^{ijkl}{\rm F}_{ij}\Gamma^{t}\theta \partial_k y^ {t} \bigr)
\eea
Using a series expansion for $ \sqrt{\det{(G_{ij}+2\pi\alpha ^\prime {\cal F}_{ij} )}}$ and a standard redefinition of the fields we get the effective Lagrangian up to the fourth order in fields (and coupling constant $\alpha = 2\pi\alpha ^\prime$):
\bea
{\it L}^4_{BI+WZ} & = & 2 i\bar \theta\dslash \theta +\frac{1}{4}{\rm F}_{ij}{\rm F}_{ji}-\frac{1}{2}\partial_i y^ {t}\partial_i y^ {t}  +  \alpha^2 \bigl(  {(\bar \theta\dslash \theta )}^2-\bar \theta \Gamma^{i}\partial_j\theta \bar \theta \Gamma^{j}\partial_i\theta \bigr) \nonumber \\
& + & \frac{\alpha^2 }{8}\bigl({\rm F}_{ij}{\rm F}^{jk}{\rm F}_{kl}{\rm F}^{li}- \frac{1}{4}({\rm F}_{ij}{\rm F}_{ji})^2 \bigr) +\frac{\alpha^2 }{4}\bigl(\partial_i y^ {t}\partial_j y^ {t}\partial_i y^ {t^\prime}\partial_j y^{t^\prime}-\frac{1}{2}(\partial_i y^ {t}\partial_i y^ {t})^2 \bigr) \nonumber \\
& - &\frac{\alpha^2 }{2}\bigl(\partial_i y^ {t}\partial_j y^ {t}{\rm F}_{ik}{\rm F}^{kj}-
\frac{1}{4}\partial_i y^ {t}\partial_i y^ {t}{\rm F}_{jk}{\rm F}^{kj}\bigr)\nonumber \\  
& - & i\alpha^2 \bigl((\bar \theta \Gamma^{i}\partial_j \theta )(\partial_i y^ {t} \partial_j y^ {t})- \frac{1}{2}(\bar \theta\dslash \theta )(\partial_i y^ {t}\partial_i y^ {t})+ \frac{1}{2}\partial_k \bar \theta \Gamma^{ijk}\partial_i y^{t} \Gamma^{t}\partial_j y^ {t^\prime} \Gamma^{t^\prime}\theta \bigr)\nonumber \\ 
&+ & i\alpha^2 \bigl(\bar \theta \Gamma^{i}\partial_j\theta{\rm F}^{jk}{\rm F}_{ki}-\frac{1}{4}\bar \theta \dslash \theta {\rm F}^{ij}{\rm F}_{ji} \bigr)\nonumber \\ 
&+& i\alpha^2\bigl(\bar \theta \partial_{i} y^ {t}\Gamma^{t} \partial_{j}\theta {\rm F}_{ij} -\frac{1}{2}\bar \theta \partial_{i} y^{t} \Gamma^{t}\Gamma^{(4)} \partial_{l}\theta {\rm F}_{jk}\eps ^{ijkl}\bigr) 
\eea

To establish the analogy between this theory and ${\cal N}=4$ super-Yang-Mills theory it is helpful to use the chiral representation for $\tilde{\gamma}^t$ (see Appendix) where the SU(4) symmetry is manifest.  Using this representation we will  introduce new notations for the scalar fields  $s^{IJ}=\frac{1}{2}(\tilde{\sigma }_t^{-1})^{IJ} y^{t} $  with SU(4) indices $I,\,J =1,2,3,4.$  The 16-component spinor $\theta $ will be represented as four  d=4  Majorana spinors,  
\bea
 \theta =\frac{{\tilde \psi}^{(I)}}{2}=\left(
\begin{array}{cc}
\lambda_{\alpha I}\\
\bar {\lambda}^{\dot{\alpha}I}
\end{array}
\right),
\eea 
 where the  extra factor $ 2 $ was introduced for convenience and the d=4 chiral projections are 
\bea
{\tilde \psi}^+_I =\frac{1}{2}(1+ \gamma^{5}){\tilde \psi}^{(I)}; \quad 
{\tilde \psi}^{-I}=\frac{1}{2}(1- \gamma^{5}){\tilde \psi}^{(I)}.\nonumber
\eea
The quartic ${\cal N}=4$ Lagrangian in terms of the new 4-dim YM variables $ (g,{\tilde \psi}^I,s^{IJ})$ with $ g^i=A^i$ is
 \bea
&&{\it L}^4_{BI+WZ} =  i\frac{1}{2}\bar{\tilde \psi}_I \dslash {\tilde \psi}^I +\frac{1}{4}{\rm F}_{ij}{\rm F}_{ji}-\frac{1}{2}\partial_i s^{IJ}\partial_i s_{IJ} \nonumber \\
& +&  \frac{\alpha^2}{16} \Bigl(  {(\bar{\tilde \psi}_I\dslash {\tilde \psi}^I )}^2-\bar{\tilde \psi}_I \gamma^{i}\partial_j{\tilde \psi}^I \bar{\tilde \psi}_J \gamma^{j}\partial_i {\tilde \psi}^J \Bigr) \nonumber \\
& + & \frac{\alpha^2 }{8}\Bigl({\rm F}_{ij}{\rm F}^{jk}{\rm F}_{kl}{\rm F}^{li}- \frac{1}{4}({\rm F}_{ij}{\rm F}_{ji})^2 \bigr) +\frac{\alpha^2 }{4}\bigl(\partial_i s^{IJ}\partial_j s_{IJ}\partial_i s^{I^\prime J^\prime}\partial_j s_{I^\prime J^\prime}-\frac{1}{2}(\partial_i s^{IJ}\partial_i s_{IJ})^2 \Bigr) \nonumber \\
& - &\frac{\alpha^2}{2}\Bigl(\partial_i s^{IJ}\partial_j s_{IJ}{\rm F}_{ik}{\rm F}^{kj}- \frac{1}{4}\partial_i s^{IJ}\partial_i s_{IJ}{\rm F}_{jk}{\rm F}^{kj}\Bigr)\nonumber \\  
& - & \frac{i\alpha^2}{4} \Bigl((\bar{\tilde \psi}_I \gamma^{i}\partial_j {\tilde \psi}^I )(\partial_i s^{IJ} \partial_j s_{IJ})- \frac{1}{2}(\bar{\tilde \psi}_I\dslash {\tilde \psi}^I )(\partial_i s^{IJ}\partial_i s_{IJ})\nonumber \\
&- &2\partial_k \bar{\tilde \psi}_I\gamma^{ijk}\partial_i s^{IJ} \partial_j s_{JK}{\tilde \psi}^K\Bigr)\nonumber \\ 
&+ & \frac{i\alpha^2}{4}\Bigl(\bar{\tilde \psi}_I \gamma^{i}\partial_j {\tilde \psi}^I{\rm F}^{jk}{\rm F}_{ki}-\frac{1}{4}\bar{\tilde \psi}_I \dslash {\tilde \psi}^I {\rm F}^{ij}{\rm F}_{ji} \Bigr)\nonumber \\ 
&+ &\frac{i\alpha^2}{2}\Bigl(\bar{\tilde \psi}_I^+  \partial_{i} s^{IJ}\gamma^{5}
\partial_{j}{\tilde \psi}_{J}^+{\rm F}_{ij}-\bar{\tilde \psi}^{I-}\partial_{i} s^{IJ}\gamma^{5}\partial_{j}{\tilde \psi}^{J-}{\rm F}_{ij}\nonumber \\
&+& \frac{1}{2}\bar{\tilde \psi}_I^+\gamma^{ijkl}\partial_{i}s^{IJ}\gamma^{5} \partial_{l}{\tilde \psi}_J^+{\rm F}_{jk}-\frac{1}{2}\bar{\tilde \psi}_I^-\gamma^{ijkl}\partial_{i} s^{IJ}\gamma^{5}\partial_{l}{\tilde \psi}_J^- {\rm F}_{jk}\Bigr).
\label{hlarg1} 
\eea
where $\gamma^{i_1\cdots i_k}\equiv \frac{1}{k!}\gamma^{[i_1}\cdots \gamma^{i_k]}. $ The parameter $\tau $ from Eq.(\ref{tau1}) is, up to linear order in the fields: 
\bea
\tau_{IJ} \approx \frac{1}{2}\dslash y^ {t}\Gamma^{t}_{IJ}-\frac{1}{4}\gamma^{ij}{\rm F}_{ij}\delta_{IJ}.
\eea
The supersymmetry  transformation  now looks like:
\bea
\delta_{susy}{\tilde \psi}^{(I)}&=&\left(
\begin{array}{cc}
-\frac{1}{2}\gamma^{ij}{\rm F}_{ij}\delta_{IJ} & 2\dslash \gamma^{5}s_{IJ}\\
 -2\dslash \gamma^{5}s^{IJ} &  -\frac{1}{2}\gamma^{ij}{\rm F}_{ij}\delta_{IJ}
\end{array}
\right)\cdot\left(
\begin{array}{cc}
\varepsilon^-_J\\
\bar{\varepsilon} ^{-J}
\end{array}
\right) \nonumber\\
\delta_{susy}g^i&=&i\bar{\tilde \psi}\gamma^{i}\varepsilon^-\nonumber\\
\delta_{susy}s_{IJ}&=&i{\varepsilon^-}_I{\tilde \psi}_J-i{\varepsilon^-}_J{\tilde \psi}_I+i{\bar\varepsilon ^-}_K\bar{\tilde \psi}_L\eps ^{IJKL}.\nonumber
\eea
Finally, using a chiral representation for $\gamma^{i}$ we get
\bea
\delta_{susy}{\tilde \psi}_{I}&=&-\frac{1}{2}\sigma^{ij}{\rm F}_{ij}{\varepsilon^-}_I- 2\sigma^{i}\partial_{i}s_{IJ} \bar{\varepsilon} ^{-J} \label{susyfc4} \\
\delta_{susy}g^i&=&-i\varepsilon^{-}_I\sigma^i\bar{\tilde \psi}_I+i{\tilde \psi}^I\sigma^i(\bar{\varepsilon}^-)^{I}\label{susyvc4}\\
\delta_{susy}s_{IJ}&=&i{\varepsilon^-}_I{\tilde \psi}_J-i{\varepsilon^-}_J{\tilde \psi}_I+i{\bar\varepsilon ^-}_K\bar{\tilde \psi}_L\eps ^{IJKL}\label{susyscc4}
\eea
which is exactly the ${\cal N}=4$ transformation given in \cite{Sohnius}.

\section{Helicity amplitudes.}

To calculate one loop corrections to the  D3-brane action we use the helicity amplitude technique (see for example the  review paper  \cite{SHform} and references therein). In the spinor helicity formalism, positive- and negative-helicity  massless spinors $\psi$ are represented as
\be
|p\pm\rangle = \psi_{\pm}(p)= \frac{1}{2}(1\pm \gamma^{5}) \psi(p)
\ee
with antisymmetric spinor products
\bea
\langle p_i-|p_j+\rangle &\equiv& \langle p_ip_j\rangle\equiv \langle ij\rangle\\
\langle p_i+|p_j-\rangle & \equiv& [ p_ip_j] \equiv [ij]\\
\langle ij\rangle[ji] &=&2 p_i\cdot p_j\equiv s_{ij}.
\eea 
Negative- and positive-helicity  polarization vectors also can be represented through spinors $|p\pm\rangle,$
\be
\varepsilon ^+_{\mu}(p;k)=\frac {\langle k-|\gamma_\mu|p-\rangle }{\sqrt 2 \langle kp\rangle }\,,\quad  \varepsilon ^-_{\mu}(p;k)=-\frac {\langle k+|\gamma_\mu|p+\rangle }{\sqrt 2 [kp]},
\ee
where $ k_\mu $ ($k^2=0$) is a ``reference'' momentum which corresponds to the particular choice of gauge for the external legs. Due to the gauge invariance of the amplitudes, the  vector  $ k $ drops out of the final expressions.
 The supersymmetry transformations (\ref{susyfc4}),(\ref{susyvc4}) and (\ref{susyscc4})  can also be rewritten in terms of bosonic states with definite helicity $g^\pm (p)=\varepsilon ^{\pm}_{\mu}(p)g^\mu$ and  spinors ${\tilde \psi}_{\pm}(p) = |p \pm \rangle $ as  \cite{susyamp}
\bea
\lbrack Q_I (p),{\tilde \psi}^{\pm}_J (k)\rbrack & = & \mp \Gamma^{\mp}(k,p) g^{\pm}(k)\delta_{IJ}\mp\Gamma^{\pm}(k,p) s^{\pm}_{IJ}\,,\label{susyf4}\\
\lbrack Q_I (p), g^{\pm}(k) \rbrack & = & \mp i\Gamma^{\pm} (k,p) {{\tilde \psi}^{\pm}}_I, \label{susyv4}\\
\lbrack Q_I(p),s^{\pm}_{JK}(k) \rbrack & = &  \pm i\Gamma^{\mp}\delta^{IJ}{{\tilde \psi}^{\pm}}_K \mp i\Gamma^{\mp}\delta^{IK}{{\tilde \psi}^{\pm}}_J \pm i \Gamma^{\pm}{{\tilde \psi}^{\pm}}_L \eps ^{IJKL}, \label{susysc4}
\eea
where $\Gamma^+(k,p)=\bar\eta[pk],\,\quad\Gamma^-(k,p)=\eta\langle pk\rangle,$
and $\eta$ is a numerical Grassmann parameter.

   Using the Lagrangian (\ref{hlarg1}) it is easy to find the tree-level helicity amplitudes. For the vector bosons we have:
 \bea
{4!}\bigl({\rm F}_{ij}{\rm F}^{jk}{\rm F}_{kl}{\rm F}^{li}- \frac{1}{4}({\rm F}_{ij}{\rm F}_{ji})^2 \bigr)&=&(t_8)_{\mu_1\nu_1\mu_2\nu_2\mu_3\nu_3\mu_4\nu_4}F_{\mu_1\nu_1}F_{\mu_2\nu_2}F_{\mu_3\nu_3}F_{\mu_4\nu_4}\nonumber \\
&=&16(t_8)_{\mu_1\nu_1\cdots \mu_4\nu_4}k_1^{\mu_1}k_2^{\mu_2}k_3^{\mu_3}k_4^{\mu_4}\epsilon_1^{\nu_1}\epsilon_2^{\nu_2}\epsilon_3^{\nu_3}\epsilon_4^{\nu_4}
\eea
where $(t_8)$ is given in eq.(9.A.19) in  \cite{Witten}
\bea
(t_8)_{\mu_1\nu_1\mu_2\nu_2\mu_3\nu_3\mu_4\nu_4} &=&\Bigl( -\frac{1}{2}\bigl( \delta_{\mu_1\mu_2}\delta_{\nu_1\nu_2}-\delta_{\mu_1\nu_2}\delta_{\nu_1\mu_2}\bigr)\bigl( \delta_{\mu_3\mu_4}\delta_{\nu_3\nu_4}-\delta_{\mu_3\nu_4}\delta_{\nu_3\mu_4}\bigr)\nonumber \\
&+&\frac{1}{2}\bigl(\delta_{\nu_1\mu_2}\delta_{\nu_2\mu_3}\delta_{\nu_3\mu_4}\delta_{\nu_4\mu_1} + \rm{ \,antisym.\,of}\,[{\mu_i\nu_i}]\bigr)\Bigr)\nonumber \\
&+&\Bigl( (1324)+ (1342)\, \rm {permutations} \Bigr)
\eea

and $k_i$ and $\epsilon_i$ are the  momentum and polarization vectors for an external vector boson leg. 

In the YM theory  with non-Abelian gauge group SU($N_c$) the  tree amplitudes are usually represented according to color decomposition \cite{SHform,Oneloop3} as a sum of partial sub-amplitudes with fixed cyclic ordering of external legs multiplied by the color factors:
\be
{\cal A}^{tree}_n(1,2,\cdots n)=\sum_{\sigma \in \frac{S_n}{Z_n}}{\rm Tr}(T^{a_{\sigma(1)}}T^{a_{\sigma(2)}}\cdots T^{a_{\sigma(n)}} )A^{tree}_n(\sigma(1),\sigma(2),\cdots \sigma(n)) 
\ee 
where $\frac{S_n}{Z_n}$ is the set of all non-cyclic permutations and $T^{a_{\sigma(n)}}$ are matrices of the fundamental representation of the SU($N_c$) color group. In the D3-brane case the gauge group is an Abelian U(1) and color factors will be simply reduced  to 1. 
 
All four-particle tree amplitudes are proportional to ${\alpha^2}$ (\ref{hlarg1}) and to simplify the notations we will let ${\alpha^2}=1.$ Due to the supersymmetric Ward identities (SWI) \cite{susyamp} the only nonzero four vector-boson tree amplitude in this case is $A_4^{tree}(g_1^-g_2^-g_3^+g_4^+)$ (with two $-$ and two $+$ helicity bosons)

\vbox{
\begin{picture}(0,0)
%diagram 1
\Vertex(85,-70) 2
\Gluon(65,-50)(105,-90) 3 8\Gluon(105,-50)(65,-90) 3 8
\end{picture}
\bea
A^{tree}_4(1^-,2^-,3^+,4^+)&=&-i\frac{4!}{8}\langle 3,\,4|\bigl({\rm F}_{ij}{\rm F}_{jk}{\rm F}_{kl}{\rm F}_{li}-\frac{1}{4}({\rm F}_{ij}{\rm F}_{ji})^2 \bigr)|1,\,2\rangle \nonumber \\
&=&-i\frac{4}{2}(t_8)_{\mu_1\nu_1\mu_2\nu_2\mu_3\nu_3\mu_4\nu_4}k_1^{\mu_1}k_2^{\mu_2}k_3^{\mu_3}k_4^{\mu_4}\epsilon_1^{\nu_1}\epsilon_2^{\nu_2}\epsilon_3^{\nu_3}\epsilon_4^{\nu_4} \nonumber \\
&=&\frac{i}{2}\frac{st{\langle 12\rangle}^4}{\langle 12\rangle\langle 23\rangle\langle 34\rangle\langle 41\rangle}
\label{gggg}
\eea}
where $s=s_{12},\,t=s_{14},\,u=s_{13}$ are Mandelstam variables. This amplitude is related to  the standard  ${\cal N}=4$  SYM four-gluon  color-ordered sub-amplitude  \cite{SHform}  by:
\bea
A^{tree}_4(1^-,2^-,3^+,4^+)&=&\frac{1}{2}st A^{tree,YM}_4(1,2,3,4).
\label{4vect}
\eea
It is important to notice that the RHS of (\ref{4vect}) is totally symmetric under external leg permutations \cite{susyamp}.

The nonzero vector-scalar  tree amplitudes are:

\vbox{
\bea
A_4^{tree}(&g^-,& s_{IJ}^-, s_{IJ}^+,g^+)= \nonumber \\
& = & -i\frac{4}{2}(t_8)_{\mu_1t\mu_2t\mu_3\nu_3\mu_4\nu_4}k_1^{\mu_1}k_2^{\mu_2}k_3^{\mu_3}k_4^{\mu_4}\epsilon_3^{\nu_3}\epsilon_4^{\nu_4}\nonumber \\
& =& \frac{i}{2}{\langle 12\rangle}{\langle 13\rangle}[ 43][ 42] \nonumber \\
& = &\frac{st\langle 12\rangle ^2 \langle 13\rangle ^2}{2\langle 12\rangle ^4}A^{tree,YM}_4(1,2,3,4).
\label{ssgg}
\eea
\begin{picture}(0,0)
%diagram 2
\Vertex(85,70) 2
\DashLine(65,50)(85,70) 4 \Gluon(85,70)(105,90) 3 5\DashLine(105,50)(85,70) 4
\Gluon(85,70)(65,90) 3 5
\put(60,95){\makebox(0,0)[cc]{$g^-$}}
\put(100,95){\makebox(0,0)[cc]{$g^+$}}
\put(60,45){\makebox(0,0)[cc]{$s_{IJ}$}}
\put(100,45){\makebox(0,0)[cc]{$s_{IJ}$}}
\end{picture}}

There are two kinds of four-scalar amplitudes:

\vbox{ 
\bea
 A_4^{tree}( &s_{IJ}^-,&s_{KL}^-, s_{IJ}^+,s_{KL}^+) = \frac{i}{2} st\label{ssss1}\\
&=& \frac{st\langle 12\rangle \langle 23\rangle \langle 34\rangle \langle 41\rangle }{2\langle 12\rangle ^4} A^{tree,YM}_4(1,2,3,4) \nonumber \\
A_4^{tree}(& s_{IJ}^-,&s_{IJ}^-, s_{IJ}^+,s_{IJ}^+)= -\frac{i}{2}s^2 \label{ssss2}\\
& = & \frac{st\langle 12\rangle ^2  \langle 34\rangle ^2  }{2\langle 12\rangle ^4} A^{tree,YM}_4(1,2,3,4). \nonumber   
\eea
\begin{picture}(0,0)
%diagram 3
\Vertex(55,105) 2
\DashLine(35,85)(75,125) 4\DashLine(75,85)(35,125) 4
\put(30,130){\makebox(0,0)[cc]{$s_{IJ}$}}
\put(70,130){\makebox(0,0)[cc]{$s_{IJ}$}}
\put(30,80){\makebox(0,0)[cc]{$s_{KL(IJ)}$}}
\put(70,80){\makebox(0,0)[cc]{$s_{KL(IJ)}$}}
\end{picture}}

The four fermion tree amplitudes  include two nonzero ones:

\vbox{
\bea
\quad \quad \quad\quad \quad A_4^{tree}(&{\tilde \psi}^-_I,&{\tilde \psi}^-_J,{\tilde \psi}^+_J,{\tilde \psi}^+_I)= -\frac{i}{2}[13][42]\langle 12\rangle ^2 \nonumber \\
&=& \frac{-\langle 12\rangle ^2  \langle 13\rangle \langle 24\rangle }{\langle 12\rangle ^4}A^{tree}_4(1,2,3,4)\label{ffff1} \\
\quad \quad \quad \quad \quad A_4^{tree}(&{\tilde \psi}^-_I,&{\tilde \psi}^-_I,{\tilde \psi}^+_I,{\tilde \psi}^+_I)=-\frac{i}{2}[12][43]\langle 12\rangle ^2 \nonumber \\
&= &\frac{-st\langle 12\rangle ^3  \langle 34\rangle }{2\langle 12\rangle ^4}A^{tree,YM}_4(1,2,3,4).\label{ffff2} 
\eea
\begin{picture}(0,0)
%diagram 4
\Vertex(75,95) 2
\Line(55,75)(95,115) \Line(95,75)(55,115) 
\put(50,120){\makebox(0,0)[cc]{${\tilde \psi_I}^-$}}
\put(90,120){\makebox(0,0)[cc]{${\tilde \psi_I}^+$}}
\put(50,70){\makebox(0,0)[cc]{${\tilde \psi_{J(I)}}^-$}}
\put(90,70){\makebox(0,0)[cc]{${\tilde \psi_{J(I)}}^+$}}
\end{picture}}

The next set of diagrams contains a mixture of two fermions and two bosons. There are five  nonzero tree amplitudes:

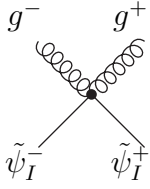
\vbox{
\bea
A_4^{tree}(&g^-,&{\tilde \psi}^-_I,{\tilde \psi}^+_I,g^+) = -(i)\frac{i}{2}[43][24]\langle 12\rangle ^2 \label{ggff}\\
& = &(i)\frac{-st\langle 12\rangle ^3 \langle 13\rangle }{2\langle 12\rangle ^4} A^{tree,YM}_4(1,2,3,4)\\ 
\nonumber\\
\nonumber
\eea
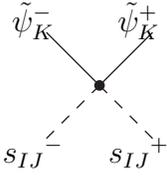
\begin{picture}(0,0)
%diagram 5
\Vertex(35,80) 2
\Line(15,60)(35,80) \Gluon(35,80)(55,100) 3 5\Line (55,60)(35,80)
\Gluon(35,80)(15,100) 3 5
\put(10,110){\makebox(0,0)[cc]{$g^-$}}
\put(50,110){\makebox(0,0)[cc]{$g^+$}}
\put(10,55){\makebox(0,0)[cc]{${\tilde \psi_{I}}^-$}}
\put(50,55){\makebox(0,0)[cc]{${\tilde \psi_{I}}^+$}}
\end{picture}}

\vbox{
\bea
\quad \quad \quad\quad A_4^{tree}(&{\tilde \psi}^-_I,& s_{JK}^-,s_{JK}^+,{\tilde \psi}^+_I) = -(i)\frac{i}{2}[13][24]\langle 12\rangle \langle 13\rangle\nonumber\\
& = &(i) \frac{st\langle 12\rangle \langle 13\rangle ^2\langle 24\rangle}{2\langle 12\rangle ^4}  A^{tree,YM}_4(1,2,3,4)\label{ffss1}\\
\quad \quad \quad\quad A_4^{tree}(&{\tilde \psi}^-_I,& s_{JK}^-,s_{IJ}^+,{\tilde \psi}^+_K) = -(i)\frac{i}{2}[14][23]\langle 12\rangle \langle 13\rangle \nonumber\\
& = &(i)\frac{st\langle 12\rangle \langle 13\rangle \langle 14\rangle\langle 23\rangle}{2\langle 12\rangle ^4} A^{tree,YM}_4(1,2,3,4)\label{ffss2}\\
\nonumber
\eea
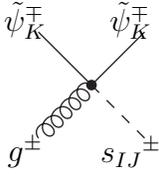
\begin{picture}(0,0)
%diagram 6
\Vertex(35,95) 2
\DashLine(15,75)(35,95) 4 \Line(35,95)(55,115)
\DashLine(55,75)(35,95) 4  \Line(35,95)(15,115) 
\put(10,120){\makebox(0,0)[cc]{${\tilde \psi_K}^-$}}
\put(50,120){\makebox(0,0)[cc]{${\tilde \psi_K}^+$}}
\put(10,70){\makebox(0,0)[cc]{${s_{IJ}}^-$}}
\put(50,70){\makebox(0,0)[cc]{${s_{IJ}}^+$}}
\end{picture}}

\vbox{
\bea
A_4^{tree}(&{\tilde \psi}^-_I,&{\tilde \psi}^-_J, s_{IJ}^+,g^+)= \label{ffgs1}\\
& = & i\frac{st\langle 12\rangle^2 \langle 13\rangle \langle 23\rangle}{2\langle 12\rangle ^4} A^{tree,YM}_4(1,2,3,4)\nonumber\\
A_4^{tree}(&g^-,& s_{IJ}^-,{\tilde \psi}^+_I,{\tilde \psi}^+_J) = \label{ffgs2}\\
& = & -i\frac{st\langle 12\rangle^2 \langle 13\rangle \langle 14\rangle}{2\langle 12\rangle ^4} A^{tree,YM}_4(1,2,3,4)\nonumber\\
\nonumber 
\eea
\begin{picture}(0,0)
%diagram 7
\Vertex(35,95) 2
\Gluon(15,75)(35,95) 3 5 \Line(35,95)(55,115)
\DashLine(55,75)(35,95) 4  \Line(35,95)(15,115) 
\put(10,120){\makebox(0,0)[cc]{${\tilde \psi_K}^{\mp}$}}
\put(50,120){\makebox(0,0)[cc]{${\tilde \psi_K}^{\mp}$}}
\put(10,70){\makebox(0,0)[cc]{${g}^{\pm}$}}
\put(50,70){\makebox(0,0)[cc]{${s_{IJ}}^{\pm}$}}
\end{picture}}

  This set of tree amplitudes has exactly the same relative factors between them  as for  ${\cal N}=4$ SUSY Yang-Mills theory given in \cite{Maxim} (extra $i$'s come from difference between  metrics signature which are  $g^{\mu\nu}=(+---)$ in \cite{Maxim} and  $g^{\mu\nu}=(-+++)$ in this paper ). The general expression connecting SYM amplitudes and the D3 brane amplitudes is:
\bea
A_4^{tree}(l_1,l_2,l_3,l_4)=\frac{st}{2}A_4^{tree,YM}(l_1,l_2,l_3,l_4)\label{d3ym}
\eea 
where $l_1,l_2,l_3,l_4$ are the momenta of the particles from N=4 SUSY multiplet.

  The main difference between the standard  ${\cal N}=4$  SYM and the D3-brane effective theory (\ref{hlarg1}) is that the first one is  finite in d=4. The following calculation shows that the D3-brane action is not one-loop finite in the flat background.

A very useful  relation for  one-loop calculations in the standard ${\cal N}=4$ SYM theory is \cite{Maxim,cutalgebra}:
\bea
\sum_{S_1,S_2\in (N=4)}&& A^{tree,YM}_4(-{\it l}_1^{S_1},1,2,{\it l}_2^{S_2})\times  A^{tree,YM}_4(-{\it l}_2^{S_2},3,4,{\it l}_1^{S_1})\nonumber \\
&=&\frac{-ist A^{tree,YM}_4(1,2,3,4)}{({\it l}_1-{\it k}_1)^2({\it l}_2-{\it k}_3)^2}
\label{treesum}
\eea
where the  sum is over all particles in   the ${\cal N}=4$  multiplet  and $1, 2, 3, 4$ stands for the  momenta of the external particles. This statement is true for any dimension and its proof can be found in  \cite{cutalgebra,bdstfe}. For our case (with relation  (\ref{d3ym})) it becomes:
\bea
\sum_{S_1,S_2\in (N=4)}&& A^{tree(D3)}_4(-{\it l}_1^{S_1},1,2,{\it l}_2^{S_2})\times  A^{tree(D3)}_4(-{\it l}_2^{S_2},3,4,{\it l}_1^{S_1})\label{D3treesum} \\
&=&\frac{-is^2}{4} st A^{tree,YM}_4(1,2,3,4)\nonumber \\
&=&\frac{-is^2}{2} A^{tree(D3)}_4(1,2,3,4)\nonumber
\eea
here $A^{tree(D3)}_4(1,2,3,4)$ is the four vector-boson D3-brane amplitude (\ref{4vect}).

 The ${\cal N}=4$ supersymmetry of the D3-brane action allows us to use a unitarity-based construction of the one loop amplitudes,  as it has done in  ${\cal N}=4$ Super-Yang-Mills theory (see  review  \cite{Onelooprev} and references therein).  In this technique, one obtains the imaginary part of the one-loop amplitude from the product of the tree amplitudes, and then reconstructs the real part up to a possible polynomial function.  The good ultraviolet behavior of  ${\cal N}=4$ super-Yang-Mills theory makes it possible to reconstruct the whole amplitude without the additive polynomial ambiguity. The one-loop amplitudes that can be reconstructed that way (``cut-constructible'')  must satisfy a certain loop-momentum  ``power-counting'' criterion  given, for example, in \cite{Onelooprev}.    

 One-loop amplitudes in the standard  ${\cal N}=4$  SYM theory obtained  by using the cut-reconstruction formalism can be expressed as:
\bea
 A^{N=4,1-loop}_4(1, 2, 3, 4)&=& ig^2s_{12}s_{23}A^{tree}_4(1, 2, 3, 4)(C_{1234}{\it I}^{1-loop}_4(s_{12},s_{23})\nonumber \\
&+&C_{3124}{\it I}^{1-loop}_4(s_{12},s_{13})+C_{2314}{\it I}^{1-loop}_4(s_{23},s_{13}))
\label{loopampl}
\eea
where $C_{1234}$ is a color factor for the non-Abelian gauge group and ${\it I}^{1-loop}_4$ is a one-loop four-point (box) integral,
\bea
{\it I}^{1-loop}_4(s_{12}s_{23})=\int \frac{d^Dp}{(2\pi)^D}\frac{1}{p^2(p-k_1)^2(p-k_1-k_2)^2(p+k_4)^2}.\label{box}
\eea
  The expression (\ref{treesum}) was used to simplify the direct product of the  tree amplitudes.
 
 The expressions (\ref{treesum})and (\ref{D3treesum}) for ${\cal N}=4$  amplitudes are correct in all dimensions and can be used \cite{Maxim,  bdstfe1} to  reconstruct  complete  massless loop amplitudes.  The D3-brane effective  Lagrangian (\ref{hlarg1}) in the  Killing gauge  does not contain three-particle vertices, hence only bubble diagrams will contribute to the one-loop corrections to the four-particle amplitudes.  In this case the expression for one-loop corrections  considerably simplifies and  can be written as:
\bea
 &&A^{D3,1-loop}_4 (1, 2, 3, 4)= \nonumber \\
&=&\frac {i\alpha^4}{4}s_{12}s_{23}A^{tree,YM}_4(1, 2, 3, 4)(s_{12}^2{\it I}^{1-loop}_2(s_{12})+ s_{13}^2{\it I}^{1-loop}_2(s_{13})+s_{14}^2{\it I}^{1-loop}_2(s_{14}))\nonumber \\
&=&\frac {i\alpha^4}{2}A^{tree(D3)}_4(1, 2, 3, 4)(s^2{\it I}^{1-loop}_2(s)+t^2{\it I}^{1-loop}_2(t)+u^2{\it I}^{1-loop}_2(u)).
\label{D3loopampl}
\eea
Here we have taken into account that $s_{12}s_{23}A^{tree,YM}_4(1, 2, 3, 4)$ from (\ref{4vect}) is symmetric under momenta permutations and color factors are gone because gauge field in (\ref{fexp}) is an Abelian  one. Instead of the box integrals (\ref{box}) expression (\ref{D3loopampl}) contains only two-point one-loop integrals:
\bea
{\it I}^{1-loop}_2(s_{12})=\int \frac{d^{4-2\epsilon}p}{(2\pi)^{4-2\epsilon}}\frac{1}{p^2(p-k_1-k_2)^2}.
\eea
This integral is ultra-violet divergent and in  dimensional reduction it has the form:
\bea
{\it I}^{1-loop}_2(s)&=& i(4\pi)^{\epsilon-2}
\frac{\Gamma (1+\epsilon)\Gamma^2(1-\epsilon)}{\epsilon (1-2\epsilon)\Gamma(1-2\epsilon)}{\bigl(\frac{-s}{\mu ^2}\bigr)}^{-\epsilon}\nonumber \\
&\sim &\frac{i}{(4\pi)^2} \frac{1}{\epsilon}\lbrack 1+\epsilon \bigl( 2-\gamma-\log \frac{-s}{\mu ^2}+ \log 4\pi+\cdots \bigr)\rbrack 
\eea
 as $\epsilon \, \rightarrow \,0.$ 
    
It is easy to find counterterms for effective D3-brane Lagrangian (\ref{hlarg1}). Their structure at the quartic level will be
\bea
(s^2+t^2+u^2)(t_8){\rm F}^4  \rightarrow (t_8)_{\mu_1\nu_1\mu_2\nu_2\mu_3\nu_3\mu_4\nu_4}\partial_{\alpha}{\rm F}_{\mu_1\nu_1}\partial_{\alpha}{\rm F}_{\mu_2\nu_2}\partial_{\beta}{\rm F}_{\mu_3\nu_3}\partial_{\beta}{\rm F}_{\mu_4\nu_4},
\eea
 or, in the  notations of  \cite{Maxim} and \cite{MarcusSt}, the counterterms  are:
\bea
T^{D3}_4 \bigl({\rm F}_{ij}{\rm F}^{jk}{\rm F}_{kl}{\rm F}^{li}- \frac{1}{4}({\rm F}_{ij}{\rm F}_{ji})^2 \bigr),
\eea
where 
\bea
T^{D3}_4=-\frac {\alpha^4}{(4\pi)^2}\frac{1}{16\epsilon}(s^2+t^2+u^2)
\label{D3loopct}
\eea

 Using the relations (\ref{gggg})-(\ref{ffgs2}) between the amplitudes with different particle content it is easy to find the corrections and the counterterms for all other four-particle interactions of ${\cal N}=4$ multiplets in this theory.

\section{Different $\kappa$-gauges}

   The Killing gauge is obviously not the only choice for fixing the  $\kappa$-gauge. In \cite{Schwarz} it was proposed to use the gauge where $\theta_1=0$ and $\theta_2=\lambda $. Then the  supersymmetry transformations will be:
 \begin{eqnarray}
\delta \bar\theta &=& \bar\epsilon + \bar\kappa (1 - \Gamma) + \xi^\mu
\partial_\mu \bar\theta \nonumber \\
\delta X^m &=& \bar\epsilon \Gamma^m \theta - \bar\kappa (1 - \Gamma)
\Gamma^m \theta + \xi^\mu \partial_\mu X^m.
\end{eqnarray}
where $\Gamma = \left(\begin{array}{cc}
0 &\zeta \\
\tilde{\zeta} & 0\end{array}\right)$ was defined in (\ref {ganzn}).

The requirement  $\delta \bar\theta_1 = 0$ and the static gauge condition $\delta X^p = 0$ determine the connection between the  $\kappa$ and $\epsilon $ parameters, so that the final supersymmetry transformations are \cite{Schwarz}:
\begin{eqnarray}
\delta_{\kappa+\varepsilon+\xi} \bar\lambda &=& \bar\epsilon_2 + \bar\epsilon_1 \zeta + \xi^\mu \partial_\mu \bar\lambda , \nonumber \\
\delta_{\kappa+\varepsilon+\xi} y^t &=& (\bar\epsilon_2 - \bar \epsilon_1 \zeta) \Gamma^t \lambda + \xi^\mu \partial_\mu y^t , \nonumber \\
\delta_{\kappa+\varepsilon+\xi} A_\mu &=& (\bar\epsilon_1 \zeta - \bar\epsilon_2) (\Gamma_\mu +
\Gamma_t \partial_\mu y^t)\lambda \nonumber \\
& &  + ({1\over 3} \bar\epsilon_2 - \bar\epsilon_1 \zeta) \Gamma_m \lambda
\bar\lambda \Gamma^m \partial_\mu \lambda + \xi^\rho \partial_\rho A_\mu +
\partial_\mu \xi^\rho A_\rho . \label{ptrans}
\end{eqnarray}
The index $m$ is a d=10  index, which includes both $\mu (p)$ and $t$ values ($t$ is a six-dimensional index for scalar fields). The induced world-volume metric in this case is:
\begin{equation}
G_{\mu\nu} = \eta_{mn} L_\mu^m L_\nu^n,\,\quad L_\mu^m  = \partial_\mu X^m - i\bar\lambda \Gamma^m \partial_\mu \lambda 
\end{equation}
and 
\begin{equation}
{\cal F}_{\mu\nu} = F_{\mu\nu} - i\bar\lambda (\Gamma_\mu + \Gamma_t \partial_\mu y^t)\partial_\nu \lambda
+ i\bar\lambda (\Gamma_\nu + \Gamma_t \partial_\nu
y^t)\partial_\mu \lambda.
\end{equation}

 The choice of $\kappa$-gauge affects only the  part of the Lagrangian that contains fermions leaving the purely bosonic terms unchanged. In the  $\theta_1=0$ gauge the fermionic part of the Lagrangian  up to the quartic order in the fields is: 

\bea
{\it L}^{(4)}_{ferm} & = & i\bar \theta\dslash \theta  + i\alpha\bar \theta \Gamma^{t}\partial_{\mu}y^ {t}\partial^{\mu}\theta +\frac{i\alpha}{2}\bigl( \bar \theta \Gamma^{\mu}\partial_\nu\theta - {\rm i}\bar \theta \Gamma^{\nu}\partial_\mu\theta \bigr){\rm F}_{\mu\nu} \nonumber \\ 
& + &\frac{\alpha^2}{2}\bigl(\bar \theta \Gamma^{t}\partial_\mu\theta \bar \theta \Gamma^{t}\partial_\mu\theta -(\bar \theta\dslash \theta)^2 \bigr) - i\alpha^2\bigl( \partial_\mu y^ {t}\partial_\nu y^ {t}\bar \theta \Gamma^{\nu}\partial_\mu\theta - \frac{1}{2}\partial_\mu y^ {t}\partial_\mu y^ {t}\bar \theta\dslash \theta \bigr) \nonumber \\ 
& + &\frac{i\alpha^2}{2}\bigl( (\bar \theta \Gamma^{\mu}\partial_\nu\theta+ \bar \theta \Gamma^{\nu}\partial_\mu\theta ){\rm F}_{\mu\lambda}{\rm F}_{\lambda\nu} -\frac{1}{2}\bar \theta\dslash \theta {\rm F}_{\mu\nu}{\rm F}_{\mu\nu}\bigr)\nonumber \\ 
& + &\frac{i\alpha^2}{2}{\rm F}_{\mu\nu}\bigl( \bar \theta \Gamma^{t}\partial_\mu y^ {t}\partial_\nu\theta-\bar \theta \Gamma^{t}\partial_\nu y^ {t}\partial_\mu\theta\bigr). 
\eea
In this gauge the Lagrangian contains the three-particles vertices along with  the quartic terms. Four-particle tree amplitudes with fermions will contain 
a set of tree diagrams where  line    \hbox{\begin{picture}(0,0) \Photon(0,3)(20,3) 3 5\end{picture} $ \quad \,\,\, $} denotes  both gauge $A_{\mu}$ and scalar $y^t$  bosons and \hbox{\begin{picture}(0,0) \Line(0,3)(20,3) \end{picture} $ \quad \,\,\, $}
 stands for the fermions $ \theta :$

 %%$$ \quad\quad $$

\vbox{
\begin{picture}(0,0)
\Vertex(35,-30) 5
\Photon(15,-10)(35,-30) 3 5 \Line(35,-30)(55,-50)
\Photon(55,-10)(35,-30) 3 5  \Line(35,-30)(15,-50)
\Vertex(95,-30) 2
\Photon(75,-10)(95,-30) 3 5 \Line(95,-30)(115,-50)
\Photon(115,-10)(95,-30) 3 5  \Line(95,-30)(75,-50)
\Vertex(155,-30) 2 \Vertex(185,-30) 2
\Photon(135,-10)(155,-30) 3 5 \Line(135,-50)(155,-30)
\Line(155,-30)(185,-30)
\Line(205,-50)(185,-30)\Photon(205,-10)(185,-30) 3 5
\put(65,-30){\makebox(0,0)[cc]{$=$}}
\put(125,-30){\makebox(0,0)[cc]{$+$}}
\put(215,-30){\makebox(0,0)[cc]{$+\, \cdots $}}
\end{picture}}

\vspace{40pt}

\pagebreak

\vbox{
\begin{picture}(0,0)
\Vertex(35,-30) 5
\Line(15,-10)(35,-30)  \Line(35,-30)(55,-50)
\Line(55,-10)(35,-30)  \Line(35,-30)(15,-50)
\Vertex(95,-30) 2
\Line(75,-10)(95,-30)  \Line(95,-30)(115,-50)
\Line(115,-10)(95,-30)  \Line(95,-30)(75,-50)
\Vertex(155,-30) 2 \Vertex(185,-30) 2
\Line(135,-10)(155,-30) \Line(135,-50)(155,-30) 
\Photon(155,-30)(185,-30) 3 5
\Line(205,-10)(185,-30)\Line(205,-50)(185,-30)
\put(215,-30){\makebox(0,0)[cc]{$+\, \cdots $}}
\put(125,-30){\makebox(0,0)[cc]{$+$}}
\put(65,-30){\makebox(0,0)[cc]{$=$}} 
\end{picture}}

\vspace{60pt}

Calculations  of these tree amplitudes show that they are the same for  both cases of the $\kappa$-gauges.  Both theories have the same one loop corrections and nontrivial counterterms in d=4.

\section{Conclusion}

The Born-Infeld type actions have received a lot of attention in the last few years.  This type of action plays a crucial role for the D3 brane theory. The rich structure of interaction terms in this theory makes it especially interesting to study them. The other interesting question that arises in this theory is a question about the form of supersymmetry transformations after the fixing of the $\kappa-$gauge. 
      
We have presented the supersymmetry transformations, Feynman rules and one-loop corrections to the $\kappa $-gauge fixed  D3 brane Born-Infeld theory up to the quartic order. We would like to stress here that the Feynman rules following from the gauge-fixed supersymmetric Born-Infeld action turned out to be rather simple at least for the 4-point one loop calculations, which has allowed us to perform the quantum  calculation in this very unusual theory.

   Our calculations show that D3 brane theory in the  flat IIB background  has a nontrivial counterterm in d=4.  We have found that the structure of the  counterterm as well as finite corrections  includes terms like $(\partial {\rm F})^4.$  We hope that our calculations may help to shed some light  
on the relations between the properties of the fundamental theory  
including the D3 branes and the quantum ${\cal N}=4$ supersymmetric YM gauge  
theory.

We also consider these calculations as a preparation for the studies  
of the more complicated theories, e.g. the D3 brane action in  
$AdS_5\times S^5$  background or, possibly, for a few interacting D3 branes with non-Abelian gauge field in the action. We have found that the use of the 
Killing gauge for $\kappa$ -symmetry combined with the helicity  
amplitudes technique has simplified our calculations significantly.
Hopefully these methods will be useful for calculations in some  
other class of interesting problems.

\vspace{40pt}

\large{\bf Acknowledgments}}

 I am grateful to R. Kallosh for the help in the statement of the problem and for support and encouragement. I am thankful to L. Dixon for very useful discussions that helped me in my calculations. I am also thankful to J. Rahmfeld and M. Perelstein with whom we have discussed many issues in the D-brane theory and helicity amplitude formalism. I would also like to thank Michael Peskin and Arkady Tseytlin for the important comments on this paper.  
 
This research was supported by the Department of Energy under grant DOE-FG05-91ER40627 and partly under contract DE-AC03-76SF00515. This work was also supported in part by the NSF grant PHY-9870115.

%%\section{Appendix}

\vspace{40pt}

{\large {\bf Appendix}}

\vspace{20pt}

We use chiral representation for d=6 dimensional $\gamma^{(6)}$ matrices following  M. Sohnius  notations \cite{Sohnius}. In this representation  $\gamma^{(6)}$ is written  as

\be
\tilde{\gamma}^t=\left(
\begin{array}{cc}
0  &  (\tilde{\sigma}_t)_{IJ}\\
(\tilde{\sigma }_t^{-1})^{IJ} &  0
\end{array}
\right)
\ee 
where matrices $(\tilde{\sigma}_t)_{IJ}$  satisfies the conditions:
\bea
\tr \tilde{\sigma }_t \tilde{\sigma }_{t^\prime}^{-1}&=& 4\delta_{t{t^\prime}}\nonumber \\
(\tilde{\sigma }_t^{-1})^{IJ}&=&-\frac{1}{2}\eps ^{IJKL}(\tilde{\sigma}_t)_{KL}
\eea

 A general  d=10 32 component complex spinor will be:
\be
\theta = \left(
\begin{array}{cc}
\lambda_{\alpha I}\\
\chi_{\alpha}^I\\
\bar{\omega}^{\dot{\alpha}}_I\\
\bar{\varphi}^{\dot{\alpha} I}
\end{array}
\right) \, ,\quad  \bar\theta =\left( {\omega}^{\alpha I} \,,{\varphi}^{\alpha}_I \,,\bar {\lambda}_{\dot{\alpha}}^I\,,\bar{\chi}_{\dot{\alpha} I}  \right )
\ee
The d=10  Majorana  and chirality conditions \cite{Sohnius}  are: 
\bea
\theta &=& C^{(10)}\bar \theta ^T\,\quad ,C^{(10)}= C^{(4)}\otimes C^{(6)}=\left( \begin{array}{cc}
-\eps_{\alpha\beta} & 0\\
 0 &-\eps^{\dot{\alpha}\dot{\beta}} 
\end{array}
\right)\otimes \left(
\begin{array}{cc}
0  &  1\\
 1 &  0
\end{array}
\right)\\
\theta &=&\frac{1}{2}(1-\Gamma^{(11)})\theta \,,\quad \Gamma^{(11)}=\gamma^5\otimes\gamma^{(7)}=i\gamma^5 \otimes \left( \begin{array}{cc}
1& 0\\
 0 &-1 
\end{array}
\right).
\eea
 They will give  the constraints on the spinors components: 
\bea
\bar{\varphi}^{\dot{\alpha} I}&=&\bar {\lambda}^{\dot{\alpha}I}=(\lambda_{\beta I})^\dag \eps^{\dot{\beta}\dot{\alpha}} \,,\quad \bar{\omega}^{\dot{\alpha}}_I=\bar{\chi}_{\dot{\alpha} I} =(\chi_{\beta}^I)^\dag \eps^{\dot{\beta}\dot{\alpha}}\\
\bar{\omega}^{\dot{\alpha}}_I&=&{\chi}_{\alpha}^I=0.
\eea
The surviving 16 components are 
\be
\theta = \left(
\begin{array}{cc}
\lambda_{\alpha I}\\
0\\
0\\
\bar{\lambda}^{\dot{\alpha} I}
\end{array}
\right);\quad 
 \bar\theta =\left( 0\,,{\lambda}^{\alpha}_I \,,
\bar {\lambda}_{\dot{\alpha}}^I \,,0\right ).
\ee

The  new  scalar fields $s_{IJ}$ with manifest SU(4) indexes $I,J=1,\cdots 4 $ are 
\bea
s_{IJ}=-\frac{1}{2}(\tilde{\sigma }_t )_{IJ} y^{t}; \,\,
s^{IJ}=\frac{1}{2}(\tilde{\sigma }_t^{-1})^{IJ} y^{t} \nonumber  
\eea
and  $ y^{t}y^{t}=s^{IJ}s_{IJ}.$

%\bibliographystyle{preprint}
%\bibliography{AdS5}

\end{document}